# Theory of the Spontaneous Emission in Photonic and Plasmonic Nanoresonators

C. Sauvan,[1, *] J.P. Hugonin,[2] I.S. Maksymov,[2] and P. Lalanne[1, †]

[1]*Laboratoire Photonique Numérique et Nanosciences,
Institut d'Optique, Université Bordeaux, CNRS, 33405 Talence, France*
[2]*Laboratoire Charles Fabry, Institut d'Optique, CNRS,
Univ. Paris Sud, 2 avenue Augustin Fresnel, 91127 Palaiseau, France*

(Dated: April 29, 2013)

We provide a self-consistent electromagnetic theory of the coupling between dipole emitters and dissipative nanoresonators. The theory that relies on the concept of quasi-normal modes with complex frequencies provides an accurate closed-form expression for the electromagnetic local density of states (LDOS) of any photonic or plasmonic resonator with strong radiation leakage, absorption and material dispersion. It represents a powerful tool to calculate and conceptualize the electromagnetic response of systems that are governed by a small number of resonance modes. We use the formalism to revisit Purcell's factor. The new formula substantially differs from the usual one; in particular, it predicts that a spectral detuning between the emitter and the resonance does not necessarily result in a Lorentzian response in the presence of dissipation. Comparisons with fully-vectorial numerical calculations for plasmonic nanoresonators made of gold nanorods evidence the high accuracy of the predictions achieved by our semi-analytical treatment.

The modification of the spontaneous decay (SD) rate of a quantum emitter induced by an electromagnetic resonance, the so-called Purcell effect [1], is one of the very fundamental effects in quantum electrodynamics. With the advent of nanotechnologies, this effect is nowadays revisited at deep-subwavelength scales with new applications in nano-optical spectroscopy [2–5], nanolasers [6], coherent generation of plasmons [7–9] or broadband single-photon sources [10, 11]. In his landmark note [1], Purcell introduced two important quantities, the quality factor $Q$ and the mode volume $V$, to quantify the maximum SD acceleration that may be achieved by coupling a quantum emitter with a cavity in the weak coupling regime, $F = 3/(4\pi^2)(\lambda_0/n)^3 Q/V$, with $\lambda_0/n$ the resonance wavelength in the material surrounding the emitter.

The Purcell factor $F$ represents the maximum acceleration for an ideal coupling between the emitter and the cavity mode, i.e., a perfect spectral, spatial and polarization matching. Once the mode field distribution is known, any deviation from perfect coupling can be calculated analytically. For instance, a spectral mismatch between the dipole frequency $\omega$ and the cavity resonance $\omega_0$ reduces the decay rate $\Gamma$ according to the usual Lorentzian lineshape [12],

$$\frac{\Gamma}{\Gamma_0} = F \frac{\omega_0^2}{\omega^2} \frac{\omega_0^2}{\omega_0^2 + 4Q^2(\omega - \omega_0)^2}, \qquad (1)$$

with $\Gamma_0$ the decay rate in the bulk material. The volume initially introduced by Purcell was a geometrical volume representing the spatial extent of the (microwave) resonator, but with the large amount of work devoted to optical microcavities in the 90's, the mode volume definition has evolved to the usual expression [12, 13]

$$V = \frac{1}{\epsilon_0 n^2} \int \epsilon(\mathbf{r})|\mathbf{E}(\mathbf{r})|^2 d^3\mathbf{r}, \qquad (2)$$

where $\epsilon_0$ is the vacuum permittivity, $\epsilon(\mathbf{r})$ is the permittivity distribution of the resonator and $\mathbf{E}$ is the cavity mode normalized such that its norm is unity at the antinode of the electric field. Although the mode volume is a purely electromagnetic quantity, its definition lacks a precise argument. Actually, $V$ is difficult to define for dissipative (non-Hermitian) systems, even for dielectric cavities where the energy dissipation simply arises from radiative leakage [13–15]. This theoretical difficulty has been recently underlined (without being solved) in the literature on metallic nanoresonators, for which absorption and dispersion have to be handled in addition to radiative leakage [16–18].

In this work, we abandon the usual description based on the electromagnetic energy in lossless and non-dispersive media. From first-principles calculations based on Maxwell's equations and the Fermi's golden rule, we propose a self-consistent classical theory for the emitter/cavity coupling. We derive a closed-form expression for the local density of states (LDOS) and a generalized Purcell formula valid for any nanocavity with radiative leakage, absorption and material dispersion, including the important case of plasmonic nanoantennas [3–9, 19]. Our findings are not marginal, as they greatly expand our current understanding. In particular, we show that a cavity mode may decelerate the total SD, *even when it is spectrally and spatially matched with the emitter*. We also evidence that a spectral mismatch does not necessarily result in a Lorentzian lineshape as in Eq. (1). In fact, Eqs. (1) and (2) appear as a specific case of the present theory, valid in the limit $Q \to \infty$, i.e., when leakage, absorption and thus dispersion can be neglected. Our theory is carefully validated by comparison with fully-vectorial numerical results obtained for plasmonic nanoantennas made of gold nanorods.

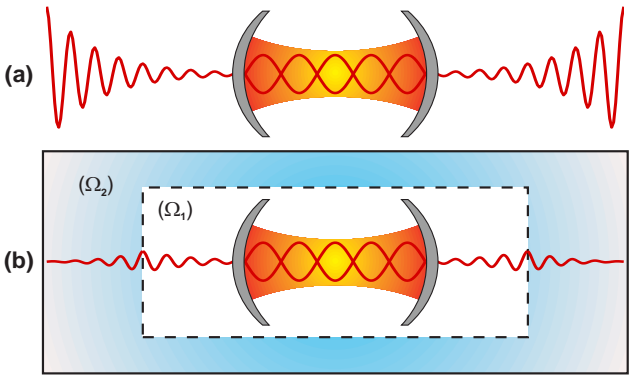

FIG. 1. Quasi-normal mode (QNM) of an open cavity. (a) QNMs are electromagnetic field distributions that satisfy Maxwell's equations for a complex frequency. The field is stationary inside the resonator and exponentially diverging outside. Because of the divergence, QNMs are not compatible with the usual expression of the mode volume in Eq. (2), which relies on the electromagnetic energy density. (b) QNM bounded by a Perfectly-Matched-Layer (PML) region shown with the bluish area ($\Omega_2$). The latter has two crucial impacts. It allows us to suppress the divergence while preserving outgoing wave boundary conditions and to calculate the mode volume by integrating over the whole domain ($\Omega_1$)∪($\Omega_2$) [20].

It is instructive to first examine what are the issues behind the definition of $V$ for dissipative cavities. The cavity modes are the electromagnetic field distributions ($\tilde{\mathbf{E}}_m, \tilde{\mathbf{H}}_m$) that are solutions of Maxwell's equations in the absence of source, $\nabla \times \tilde{\mathbf{E}}_m = i\tilde{\omega}_m \mu \tilde{\mathbf{H}}_m$ and $\nabla \times \tilde{\mathbf{H}}_m = -i\tilde{\omega}_m \epsilon(\mathbf{r}) \tilde{\mathbf{E}}_m$, and that satisfy outgoing wave boundary conditions (the Sommerfeld radiation condition as $|\mathbf{r}| \to \infty$). The tilde is related to modal quantities hereafter. Because the energy leaks out or is absorbed, the modes possess a finite lifetime $\tau_m$. The eigenfrequency $\tilde{\omega}_m$ is thus complex with $\text{Im}(\tilde{\omega}_m) = -1/\tau_m$. Consistently with the literature on open systems for which the time-evolution operator is not Hermitian [13, 14], these eigenmodes will be referred to as quasi-normal modes (QNMs), rather than normal modes, to emphasize that they are modes of a non-conservative system.

The definition of the quality factor does not raise any issue, $Q = -\frac{\text{Re}(\tilde{\omega}_m)}{2\text{Im}(\tilde{\omega}_m)}$, an expression that is related to energy balance arguments [21]. However, a problem arises with the definition of $V$. Actually, as $\text{Im}(\tilde{\omega}_m) < 0$, electromagnetic fields are amplified as they propagate (wavevectors become complex with a negative imaginary part). Thus, the QNM field diverges exponentially as $|\mathbf{r}| \to \infty$, see Fig. 1(a), and indeed, the volume integral of the intensity in Eq. (2) is also *exponentially* diverging. This problem is well known since the early studies on optical microcavities [12–15]. It has been by-passed by calculating $V$ for empirically related field distributions defined by Born-von Kármán periodic boundary conditions [12, 14]. For high-$Q$ cavities, the field that leaks in the clads is small compared to the field inside the cavity and the difference between the volumes of the empiric "mode" and of the actual QNM is negligible [12, 15]. For plasmonic nanocavities whose $Q$'s rarely exceed a few tens, the definition of $V$ is even more critical as was recently underlined [17, 18]. Not only the trick used for high-$Q$ dielectric cavities becomes largely unsubstantiated, but in addition material dispersion and Ohmic losses have to be correctly handled. The lack of a sound theoretical framework is all the more detrimental as the coupling between a quantum emitter and a plasmonic resonance is at the heart of important new paradigms, such as the stimulated generation of plasmons in spatial region much smaller than the wavelength [7–9] or the loss compensation in metamaterials [22].

We consider the general problem of an emitter located at $\mathbf{r} = \mathbf{r}_0$ in the vicinity of absorptive, dispersive and anisotropic nanostructures with open boundaries. The system is characterized by the position- and frequency-dependent permittivity and permeability tensors $\epsilon(\mathbf{r},\omega)$ and $\mu(\mathbf{r},\omega)$. We assume that the materials are reciprocal, $\epsilon = \epsilon^T$ and $\mu = \mu^T$, where the superscript denotes matrix transposition. In the weak-coupling regime, the SD rate $\Gamma$ can be derived from Fermi's golden rule and the electric Green tensor [23] (the electromagnetic response of the environment). For an electric-dipole transition at the frequency $\omega$ with a dipole moment $\mathbf{p}$, $\Gamma$ takes the form

$$\Gamma = \frac{2}{\hbar}\text{Im}[\mathbf{p}^* \cdot \mathbf{E}(\mathbf{r}_0)], \qquad (3)$$

where $\mathbf{E}$ is the total electric field that satisfies Maxwell's equations in the presence of the dipole, $\nabla \times \mathbf{E} = i\omega\mu(\mathbf{r},\omega)\mathbf{H}$ and $\nabla \times \mathbf{H} = -i\omega\epsilon(\mathbf{r},\omega)\mathbf{E} - i\omega\mathbf{p}\delta(\mathbf{r}-\mathbf{r}_0)$, with $\delta$ the Dirac distribution.

We now make the sole assumption of this work by considering that the electromagnetic field ($\mathbf{E}, \mathbf{H}$) radiated by the dipole can be expanded onto a small set of $M$ QNMs,

$$\mathbf{E}(\mathbf{r},\omega) \approx \sum_{m=1}^{M} \alpha_m(\omega)\tilde{\mathbf{E}}_m(\mathbf{r}), \qquad (4)$$

where $\alpha_m$ are complex coefficients to be determined. A similar expression with the same $\alpha_m$'s holds for the magnetic field. The number of QNMs is determined by increasing $M$ until convergence is reached ($M = 1$ in Fig. 2 and $M = 2$ in Fig. 3). If the system supports degenerate modes, they have to be all included in the expansion. The validity of Eq. (4) is questionable only when the expansion onto a set of $M$ discrete modes neglects important decay channels [17]. For instance, an emitter located outside the cavity in the evanescent field is only weakly coupled to the resonance and it will mostly decay in the free-space continuum [20]. Hereafter we first show that the $\alpha_m$'s can be easily calculated *without any approximation* by solving a linear system. In a second step, we provide an approximate analytical expression for $\alpha_m$ from which we define a generalized Purcell formula for absorptive and dispersive nanocavities.

By applying the unconjugated form of Lorentz reciprocity [20] to the total field $(\mathbf{E}, \mathbf{H})$ created by the dipole at the frequency $\omega$ and to the $n^{th}$ mode $(\tilde{\mathbf{E}}_n, \tilde{\mathbf{H}}_n)$, we obtain $\int \{\mathbf{E} \cdot [\omega\epsilon(\omega) - \tilde{\omega}_n\epsilon(\tilde{\omega}_n)]\tilde{\mathbf{E}}_n - \mathbf{H} \cdot [\omega\mu(\omega) - \tilde{\omega}_n\mu(\tilde{\omega}_n)]\tilde{\mathbf{H}}_n\} d^3\mathbf{r} = -\omega\mathbf{p}\cdot\tilde{\mathbf{E}}_n(\mathbf{r}_0)$. We then use the modal expansion of Eq. (4) and obtain a linear system of $M$ equations, $\sum_m B_{nm}(\omega)\alpha_m(\omega) = -\omega\mathbf{p}\cdot\tilde{\mathbf{E}}_n(\mathbf{r}_0)$, where the unknowns are the $\alpha_m$'s and $B_{nm}(\omega) = \int \{\tilde{\mathbf{E}}_m \cdot [\omega\epsilon(\omega) - \tilde{\omega}_n\epsilon(\tilde{\omega}_n)]\tilde{\mathbf{E}}_n - \tilde{\mathbf{H}}_m \cdot [\omega\mu(\omega) - \tilde{\omega}_n\mu(\tilde{\omega}_n)]\tilde{\mathbf{H}}_n\} d^3\mathbf{r}$.

Let us first examine these equations for non-dispersive materials. In this "simple" specific case, $\epsilon$ and $\mu$ are frequency-independent and the coefficients $B_{nm}$ become $B_{nm}(\omega) = (\omega - \tilde{\omega}_n) \int (\tilde{\mathbf{E}}_m \cdot \epsilon\tilde{\mathbf{E}}_n - \tilde{\mathbf{H}}_m \cdot \mu\tilde{\mathbf{H}}_n) d^3\mathbf{r}$. By using the orthogonality property of the QNMs of non-dispersive systems [20], we get $B_{nm}(\omega) = 0$ for $n \neq m$ and $B_{nn}(\omega) = (\omega - \tilde{\omega}_n) \int (\tilde{\mathbf{E}}_n \cdot \epsilon\tilde{\mathbf{E}}_n - \tilde{\mathbf{H}}_n \cdot \mu\tilde{\mathbf{H}}_n) d^3\mathbf{r}$. Therefore, in the absence of dispersion, the linear system of equations is diagonal and we trivially obtain $\alpha_n(\omega) = -\omega\mathbf{p} \cdot \tilde{\mathbf{E}}_n(\mathbf{r}_0) / [(\omega - \tilde{\omega}_n) \int (\tilde{\mathbf{E}}_n \cdot \epsilon\tilde{\mathbf{E}}_n - \tilde{\mathbf{H}}_n \cdot \mu\tilde{\mathbf{H}}_n) d^3\mathbf{r}]$. Note that $\alpha_n$ has a pole for $\omega = \tilde{\omega}_n$; the system resonates whenever the exciting frequency $\omega$ is close to one of the eigenfrequencies $\tilde{\omega}_n$.

In the general case of dispersive media, the QNMs are not orthogonal and the off-diagonal coefficients $B_{nm}(\omega)$ are not equal to zero. One has thus to solve a small linear system of $M \times M$ equations. To guarantee a safe and accurate numerical implementation, we note that $B_{nm}(\omega)$ is null for $\omega = \tilde{\omega}_m$ and for any $n$ or $m$ as shown in [20], and we write $B_{nm}(\omega) = (\omega - \tilde{\omega}_m) A_{nm}(\omega)$ with $A_{nm}(\tilde{\omega}_m) \neq 0$. The linear system of equations can thus be rewritten as

$$\sum_m A_{nm}(\omega) x_m(\omega) = -\omega\mathbf{p}\cdot\tilde{\mathbf{E}}_n(\mathbf{r}_0), \qquad (5)$$

where the unknowns are now $x_m(\omega) = (\omega - \tilde{\omega}_m)\alpha_m(\omega)$ and the coefficients $A_{nm}(\omega)$ are given by

$$A_{nm}(\omega) = \frac{1}{\omega - \tilde{\omega}_m} \int \{\tilde{\mathbf{E}}_m \cdot [\omega\epsilon(\omega) - \tilde{\omega}_n\epsilon(\tilde{\omega}_n)]\tilde{\mathbf{E}}_n - \tilde{\mathbf{H}}_m \cdot [\omega\mu(\omega) - \tilde{\omega}_n\mu(\tilde{\omega}_n)]\tilde{\mathbf{H}}_n\} d^3\mathbf{r}, \qquad (6)$$

with $A_{nm}(\tilde{\omega}_n) = 0$ for $n \neq m$ and $A_{nn}(\tilde{\omega}_n) = \int [\tilde{\mathbf{E}}_n \cdot \frac{\partial(\omega\epsilon)}{\partial\omega}\tilde{\mathbf{E}}_n - \tilde{\mathbf{H}}_n \cdot \frac{\partial(\omega\mu)}{\partial\omega}\tilde{\mathbf{H}}_n] d^3\mathbf{r}$. In this form, the system of equations is not singular and it can be easily solved for the $x_n$'s. We get round the difficulty associated to the divergence of the QNMs as $|\mathbf{r}| \to \infty$ by considering the field calculated in Perfectly-Matched-Layers surrounding the resonator (see Fig. 1 and [20]). Then the total decay rate is obtained from Eqs. (3) and (4), $\Gamma = \frac{2}{\hbar}\text{Im}[\sum_m \alpha_m \mathbf{p}^* \cdot \tilde{\mathbf{E}}_m(\mathbf{r}_0)]$. Equations (5)-(6) constitute the major result of this work, together with the approach developed for calculating the $A_{nm}$'s [20]; they form an accurate and efficient tool to calculate the SD rate of a quantum emitter placed in a complex nanocavity.

For dispersive materials, the linear system is not diagonal and it is not possible in general to derive a closed-form expression for the SD rate. However, we can still show that $\alpha_n(\omega)$ has a pole and use this property to derive an approximate analytical expression. For $\omega = \tilde{\omega}_n$, since $A_{nm}(\tilde{\omega}_n) = 0$ for $n \neq m$, the $n^{th}$ line of the system in Eq. (5) simply becomes $A_{nn}(\tilde{\omega}_n)x_n(\tilde{\omega}_n) = -\tilde{\omega}_n\mathbf{p}\cdot\tilde{\mathbf{E}}_n(\mathbf{r}_0)$. Since $x_n(\omega) = (\omega - \tilde{\omega}_n)\alpha_n(\omega)$, we obtain that $\alpha_n(\omega)$ has a pole for $\omega = \tilde{\omega}_n$, whose residue is given by $-\tilde{\omega}_n\mathbf{p}\cdot\tilde{\mathbf{E}}_n(\mathbf{r}_0)/A_{nn}(\tilde{\omega}_n)$. Therefore, we can write

$$\alpha_n(\omega) = \frac{-\omega\mathbf{p}\cdot\tilde{\mathbf{E}}_n(\mathbf{r}_0)}{(\omega - \tilde{\omega}_n)\int[\tilde{\mathbf{E}}_n \cdot \frac{\partial(\omega\epsilon)}{\partial\omega}\tilde{\mathbf{E}}_n - \tilde{\mathbf{H}}_n \cdot \frac{\partial(\omega\mu)}{\partial\omega}\tilde{\mathbf{H}}_n]d^3\mathbf{r}} + f_n(\omega), \qquad (7)$$

where $f_n(\omega)$ is a non-resonant background that is negligible for $\omega \approx \tilde{\omega}_n$. We thus obtain an approximate closed-form expression for $\alpha_n$ valid in the vicinity of $\tilde{\omega}_n$.

For a dipole essentially coupled to a single resonance $\tilde{\mathbf{E}}$ (an important case in practice [3, 4, 19]), the SD rate is $\Gamma = \frac{2}{\hbar}\text{Im}[\alpha\mathbf{p}^* \cdot \tilde{\mathbf{E}}(\mathbf{r}_0)]$. After normalizing by the SD rate in a bulk material with a refractive index $n$, $\Gamma_0 = \frac{\omega^3|p|^2 n}{3\pi\epsilon_0\hbar c^3}$, simple derivations lead to

$$\frac{\Gamma}{\Gamma_0} = F\frac{\omega_0^2}{\omega^2}\frac{\omega_0^2}{\omega_0^2 + 4Q^2(\omega - \omega_0)^2}\left[1 + 2Q\frac{\omega - \omega_0}{\omega_0}\frac{\text{Im}(V)}{\text{Re}(V)}\right], \qquad (8)$$

with $\omega_0 = \text{Re}(\tilde{\omega})$. In Eq. (8), $F$ and $V$ are the generalized Purcell factor and mode volume,

$$V = \frac{\int[\tilde{\mathbf{E}} \cdot \frac{\partial(\omega\epsilon)}{\partial\omega}\tilde{\mathbf{E}} - \tilde{\mathbf{H}} \cdot \frac{\partial(\omega\mu)}{\partial\omega}\tilde{\mathbf{H}}]d^3\mathbf{r}}{2\epsilon_0 n^2[\tilde{\mathbf{E}}(\mathbf{r}_0)\cdot\mathbf{u}]^2}, \qquad (9)$$

$$F = \frac{3}{4\pi^2}\left(\frac{\lambda_0}{n}\right)^3\text{Re}\left(\frac{Q}{V}\right). \qquad (10)$$

Note that we have considered a linearly-polarized dipole $\mathbf{p} = p\mathbf{u}$, with $\mathbf{u}$ a unit vector. The volume integral in Eq. (9) extends over the whole domain $(\Omega_1) \cup (\Omega_2)$ and does not diverge despite the QNM divergence, see Fig. 1 and [20]. The generalized Purcell factor in Eq. (10) takes exactly the same form as the usual factor introduced by Purcell, except that $V$ is now a complex quantity, whose real and imaginary parts impact the SD rate on- and off-resonance. The mode volume in Eq. (9) explicitly considers material dispersion as evidenced by the derivatives $\frac{\partial(\omega\epsilon)}{\partial\omega}$ and $\frac{\partial(\omega\mu)}{\partial\omega}$ taken at the complex frequency $\tilde{\omega}$. We emphasize that Eqs. (8)-(10) are valid for any dissipative system, even with large absorption.

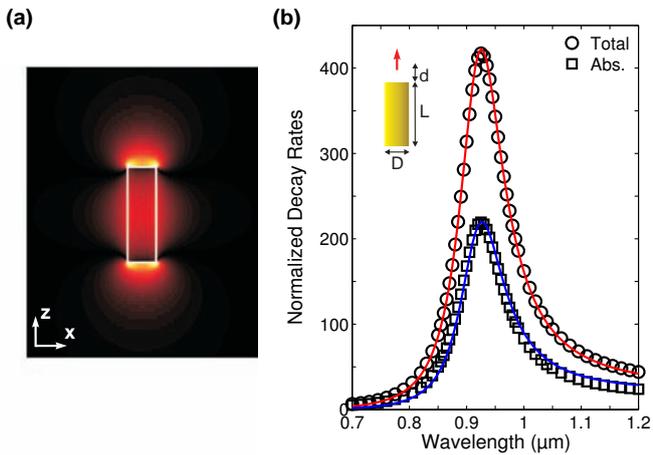

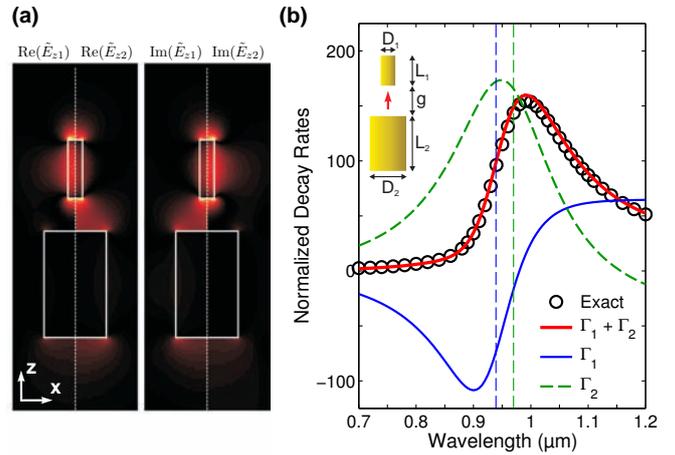

FIG. 2. Single metallic nanorod. A gold cylinder (diameter $D = 30$ nm, length $L = 100$ nm) is embedded in a host medium of refractive index $n = 1.5$. (a) Electric field distribution $|\tilde{E}_z|$ of the dipole-like QNM with a complex frequency $2\pi c/\tilde{\omega} = 920 + 47i$ nm. (b) Normalized decay-rate spectrum for an on-axis dipole oriented parallel to the nanorod (red arrow) and located at $d = 10$ nm. Circles and squares are fully-vectorial data for the total ($\Gamma$) and non-radiative ($\Gamma_{NR}$) decay rates. The solid curves are obtained from Eqs. (8) and (11) for the QNM shown in (a).

FIG. 3. Metallic gap resonator. Two gold cylinders (diameters $D_1 = 20$ nm and $D_2 = 85$ nm, lengths $L_1 = 80$ nm and $L_2 = 145$ nm) are separated by a small gap $g = 45$ nm and embedded in a host medium of refractive index $n = 1.5$. (a) Electric field distributions of the two dominant QNMs with complex frequencies $2\pi c/\tilde{\omega}_1 = 933 + 77i$ nm and $2\pi c/\tilde{\omega}_2 = 958 + 108i$ nm. The left (right) panel compares the real (imaginary) parts of $\tilde{E}_{z1}$ and $\tilde{E}_{z2}$. (b) Normalized decay-rate spectrum for an on-axis $z$-oriented dipole located in the center of the gap (red arrow). The vertical dashed lines represent $\text{Re}(\tilde{\omega}_1)$ and $\text{Re}(\tilde{\omega}_2)$. The independent contributions $\Gamma_1$ and $\Gamma_2$ of the two modes in (a) are calculated with Eq. (8) and shown by the solid blue and dashed green curves. Their sum (bold red curve) predicts a total decay rate in quantitative agreement with fully-vectorial calculations (black circles).

For a conservative (closed and lossless) cavity, the QNM field is real and Eqs. (8)-(10) reduce to the usual expressions, which then appear as valid in the limit of large $Q$'s. Energy dissipation results in the appearance of an imaginary part in the QNM field, and thus in $V$. A comparison between Eqs. (1) and (8) evidences this stringent difference between a conservative and a dissipative (open and/or lossy) system: in general, the SD rate of spectrally-detuned dipoles is not described by a Lorentzian lineshape, especially if $\text{Im}(V)/\text{Re}(V) \geqslant 1/Q$.

The present formalism is not only capable of accurately predicting the total decay rate $\Gamma$ but also the radiative ($\Gamma_R$) and non-radiative ($\Gamma_{NR}$) decays. $\Gamma_{NR}$ is due to absorption and is predicted by integrating the electric-field-intensity over the lossy region,

$$\Gamma_{NR} = \frac{2}{\hbar}|\alpha(\omega)|^2 \int \text{Im}(\epsilon)|\tilde{\mathbf{E}}(\mathbf{r})|^2 d^3\mathbf{r}, \qquad (11)$$

and the radiative decay rate is simply obtained by energy conservation, $\Gamma_R = \Gamma - \Gamma_{NR}$.

We validate the present theory by carefully testing its predictions against fully-vectorial calculations for plasmonic nanoantennas made of metallic nanorods. We first consider a single gold nanorod, see Fig. 2, which has received considerable attention to control the spontaneous emission by plasmonic modes [3, 4, 19, 24]. In the spectral range of interest, a single QNM is dominant, namely the dipole-like mode of the nanorod shown in Fig. 2(a). We have used this mode in Eqs. (8) and (11) to calculate the total ($\Gamma$) and non-radiative ($\Gamma_{NR}$) decay rates for an on-axis dipole oriented parallel to the nanorod and located at a distance $d = 10$ nm from the metal surface. As shown in Fig. 2(b), the predictions of our theory (solid curves) are in excellent agreement with fully-vectorial numerical data (circles and squares) calculated with the aperiodic Fourier Modal Method (a-FMM) [25] implemented in cylindrical coordinates [26]. To calculate QNMs, one needs an analytical continuation of the gold permittivity for complex frequencies. We have used a Drude model that fits the tabulated data in [27], $\epsilon = 1 - \omega_p^2/(\omega^2 + i\omega\gamma)$ with $\omega_p = 1.26 \times 10^{16} s^{-1}$ and $\gamma = 1.41 \times 10^{14} s^{-1}$. The mode volume calculated with Eq. (9) is mostly real and positive, $V = (5 - 0.4i)\lambda^3/10^4$.

We now evidence the accuracy of the theory with a more complex example where the emission lineshape is far from being Lorentzian. We consider a gap antenna made of two closely placed gold nanorods [4, 28] and calculate the total SD rate $\Gamma$ for an on-axis dipole located in the center of the gap, see Fig. 3. Fully-vectorial calculations show a single asymmetric resonance (circles), whereas, in this spectral range, two QNMs are spatially matched with the dipole, see Fig. 3(a). With the present formalism, $\Gamma$ is given by the sum of the independent contributions $\Gamma_1$ and $\Gamma_2$ of the two QNMs, which are calculated with Eq. (8) (solid and dashed curves). The sum (bold red curve) predicts a total SD rate in quantitative agreement with fully-vectorial calculations. We emphasize that the contribution $\Gamma_1$ to the total decay is mostly negative, *even if the corresponding mode is spectrally and spatially matched with the dipole*. This effect related to energy dissipation is accurately predicted by the present theory; it is formalized by a complex mode volume, $V_1 = (-3 - 7i)\lambda^3/10^4$. In contrast, the contribution $\Gamma_2$ is quasi-Lorentzian with $V_2 = (4 + 3i)\lambda^3/10^4$.

In conclusion, we have revisited the usual Purcell factor by providing a self-consistent electromagnetic treatment of the LDOS of dissipative nanoresonators. The latter possess QNMs with a finite lifetime that are not orthogonal in the sense of the energy. Consequently, the contribution of a QNM may decelerate the total decay, even when it is spectrally and spatially matched with the source. This unexpected effect is due to the presence of dissipation and is formalized by the signed term $\text{Re}(1/V)$ in the generalized Purcell factor of Eq. (10). Moreover, for emitter frequencies detuned from the cavity resonance, the response of the system can be non-Lorentzian. This second effect is taken into account by affecting a complex value to the mode volume, see Eq. (8). This choice may appear motivated by mathematical rather than physical considerations, but we emphasize that the new definition of $V$ is fully consistent with the usual one in Eq. (2) that is valid in the limit $Q \to \infty$. The present theory is a powerful tool since it provides highly-accurate semi-analytical predictions for most ultrasmall resonators of current interest in nanophotonics, including situations with radiation leakage, absorption and dispersion. Once a few dominant modes have been calculated, any variation of the dipole frequency, location or orientation is treated analytically, in contrast to full numerical methods. We therefore believe that the present theory may be useful to engineer further quantum effects, such as strong coupling in weakly damped cavities [29], lasing with ultrasmall plasmonic modes [7–9] or disordered systems [30], or superradiance effects in complex media [31]. Furthermore, we expect that the formalism can be carried out of nanophotonics back to classical antenna theory and used to solve the relevant problem of radiated power enhancement of electrically small antennas in realistic environments [32].

---

# Supplementary Material
# Theory of the Spontaneous Emission in Photonic and Plasmonic Nanoresonators

C. Sauvan[1], J.-P. Hugonin[2], I.S. Maksymov[2] and P. Lalanne[1]

[1]Laboratoire Photonique, Numérique et Nanosciences, Institut d'Optique, Univ Bordeaux 1, CNRS, 33405 Talence Cedex, France.
[2]Laboratoire Charles Fabry, Institut d'Optique, CNRS, Univ Paris-Sud, 2 avenue Augustin Fresnel, 91127 Palaiseau Cedex, France.

We provide hereafter some technical elements concerning the analytical derivations and the numerical calculations presented in the main text:

(1) The unconjugated form of Lorentz reciprocity theorem that is used to derive Eq. (5) in the main text is first presented in Section 1. The derivation is classical and is presented here for the sake of completeness.
(2) Section 2 contains an original and important derivation; it shows that the mode volume defined by Eq. (9) (main text) can be calculated with any Perfectly-Matched-Layer (PML). This is possible because PMLs are complex coordinate transforms under which important electromagnetic quantities remain invariant, see Annex 2 in [1]. The field inside the PML is usually considered as useless except for numerical considerations (e.g. checking the damping of the field), but we show that it has to be included to evaluate the mode volume. The use of PML as more than a numerical tool is, to our knowledge, unique in this case.
(3) Section 3 presents additional numerical tests. Like in the main text, the predictions of the formalism are found to be highly accurate for coupled photonic-crystal microcavities and plasmonic nanocylinder cavities. For the sake of completeness, Section 3 also includes a detailed analysis of the limitations of the proposed theory, when important decay channels different from the coupling to the resonator are neglected.

## 1. Unconjugated form of Lorentz reciprocity theorem for absorbing and dispersive media

We start by the classical derivation of a general form of Lorentz reciprocity theorem [2], namely the unconjugated form that is valid for absorbing and dispersive media. Reciprocity is then used (1) to derive the linear system of equations satisfied by the mode amplitudes $\alpha_m$ [see Eq. (5) in the main text], (2) to demonstrate that the quasi-normal modes (QNMs) of non-dispersive systems are orthogonal and (3) to show that $B_{nm}(\tilde{\omega}_m) = 0$, an important property that can be seen as a "weak" orthogonality condition for dispersive systems.

The system is characterized by the position- and frequency-dependent permittivity and permeability tensors $\boldsymbol{\varepsilon}(\mathbf{r}, \omega)$ and $\boldsymbol{\mu}(\mathbf{r}, \omega)$. The sole assumption consists in considering reciprocal materials, $\boldsymbol{\mu} = \boldsymbol{\mu}^T$ and $\boldsymbol{\varepsilon} = \boldsymbol{\varepsilon}^T$, where the superscript denotes matrix transposition. The electromagnetic field (**E**,**H**) radiated by a current source density **j** at the frequency $\omega$ is given by the two curl Maxwell's equations,

$$\nabla \times \mathbf{E} = i\omega \boldsymbol{\mu}(\mathbf{r}, \omega)\mathbf{H} \text{ and } \nabla \times \mathbf{H} = -i\omega \boldsymbol{\varepsilon}(\mathbf{r}, \omega)\mathbf{E} + \mathbf{j}. \tag{S1}$$

Lorentz reciprocity theorem relates two different solutions of Maxwell's equations, ($\mathbf{E}_1$, $\mathbf{H}_1$, $\omega_1$, $\mathbf{j}_1$) and ($\mathbf{E}_2$, $\mathbf{H}_2$, $\omega_2$, $\mathbf{j}_2$) labelled by the index 1 and 2. It is derived by applying the divergence theorem to the vector $\mathbf{E}_2 \times \mathbf{H}_1 - \mathbf{E}_1 \times \mathbf{H}_2$ and by using Eq. (S1),

$$\iint_\Sigma (\mathbf{E}_2 \times \mathbf{H}_1 - \mathbf{E}_1 \times \mathbf{H}_2)\cdot d\mathbf{S} = i\iiint_\Omega \{\mathbf{E}_1 \cdot [\omega_1 \boldsymbol{\varepsilon}(\omega_1) - \omega_2 \boldsymbol{\varepsilon}(\omega_2)]\mathbf{E}_2 - \mathbf{H}_1 \cdot [\omega_1 \boldsymbol{\mu}(\omega_1) - \omega_2 \boldsymbol{\mu}(\omega_2)]\mathbf{H}_2\} d^3\mathbf{r}$$
$$- \iiint_\Omega (\mathbf{j}_1 \cdot \mathbf{E}_2 - \mathbf{j}_2 \cdot \mathbf{E}_1) d^3\mathbf{r}, \tag{S2}$$

where $\Sigma$ is an arbitrary closed surface defining a volume $\Omega$.

In the main text, we use the general form of Lorentz reciprocity theorem given by Eq. (S2) and take as the first solution of Maxwell's equations the field radiated by a point dipole $\mathbf{p}\delta(\mathbf{r} - \mathbf{r}_0)$ at the frequency $\omega$, $E_1 = E$, $H_1 = H$, $\omega_1 = \omega$ and $j_1 = -i\omega p\delta(r - r_0)$, and as the second solution the $n^{th}$ QNM of the resonator, $E_2 = \tilde{E}_n$, $H_2 = \tilde{H}_n$, $\omega_2 = \tilde{\omega}_n$ and $j_2 = 0$. We consider a closed surface $\Sigma$ that includes both domains $\Omega_1$ and $\Omega_2$ (see Fig. 1 in the main text). The presence of the Perfectly-Matched-Layer (PML) in $\Omega_2$ changes the infinite volume with outgoing wave boundary conditions to a finite closed volume *with preserved outgoing wave boundary conditions and vanishing field components at the boundaries*, see details in [1] and in the next Section on the mode-volume calculation. Because of the vanishing fields, the surface integral term is null.

Lorentz reciprocity theorem can also be used to show that the QNMs of non-dispersive systems satisfy an unconjugated orthogonality relation. For that purpose, we consider Eq. (S2) with a frequency-independent permittivity $\varepsilon$ and permeability $\mu$, taking as the first solution of Maxwell's equations the $m^{th}$ QNM, $E_1 = \tilde{E}_m$, $H_1 = \tilde{H}_m$, $\omega_1 = \tilde{\omega}_m$ and $j_1 = 0$, and as the second solution the $n^{th}$ QNM, $E_2 = \tilde{E}_n$, $H_2 = \tilde{H}_n$, $\omega_2 = \tilde{\omega}_n$ and $j_2 = 0$. Using the same domain $\Omega = \Omega_1 \cup \Omega_2$ as before, Eq. (S2) leads to

$$(\tilde{\omega}_m - \tilde{\omega}_n)\iiint_\Omega (\tilde{E}_m \cdot \varepsilon \tilde{E}_n - \tilde{H}_m \cdot \mu \tilde{H}_n) d^3r = 0. \tag{S3}$$

Therefore, for $n \neq m$, the volume integral in Eq. (S3) is equal to zero, showing that the QNMs of non-dispersive systems are orthogonal.

Finally, one may apply Eq. (S2) to the same QNMs but in the general case of dispersive materials to obtain

$$\iiint_\Omega \{\tilde{E}_m \cdot [\tilde{\omega}_m \varepsilon(\tilde{\omega}_m) - \tilde{\omega}_n \varepsilon(\tilde{\omega}_n)]\tilde{E}_n - \tilde{H}_m \cdot [\tilde{\omega}_m \mu(\tilde{\omega}_m) - \tilde{\omega}_n \mu(\tilde{\omega}_n)]\tilde{H}_n \} d^3r = 0, \tag{S4}$$

which may be rewritten as $B_{nm}(\tilde{\omega}_m) = 0$.

## 2. Mode volume calculation

This Section explains how one may calculate in practice the complex mode volume defined by Eq. (9) in the main text. We will only consider the numerator integral where we have dropped the tilde for simplicity,

$$N = \iiint (E \cdot \partial(\omega\varepsilon)/\partial\omega E - H \cdot \partial(\omega\mu)/\partial\omega H) d^3r. \tag{S5}$$

In brief, the integral is performed in the numerical space thanks to a Perfectly-Matched-Layer (PML). The latter is a classical numerical tool, often used in computational electrodynamics, which consists in a renormalization of the permittivity and permeability tensors; it is nothing else than a complex coordinate transform in Maxwell's equations [3]. Figure 1(b) in the main text shows the actual numerical space, which is formed by a portion of the real space ($\Omega_1$) surrounded by a PML ($\Omega_2$). The PML illustrated with the bluish area in Fig. 1(b) serves two purposes. First it maintains the outgoing wave boundary conditions so that the field computed in $\Omega_1$ is the actual field of the QNM. Secondly, provided that the complex PML coefficient is strong enough, it forces the QNM field to fall off exponentially in the PML and to vanish at the outer boundaries of the numerical space. We now provide a demonstration that N defined by Eq. (S5) is an invariant of the coordinate transform that can be calculated with any PML.

Hereafter, we use an orthogonal cartesian coordinate system (x,y,z) for the real space. A PML is fully specified by the complex coordinate transform [3,4]

$$\hat{X} = X(x), \hat{Y} = Y(y), \hat{Z} = Z(z), \tag{S6}$$

where X(x), Y(y) and Z(z) are complex transformations from the complex plane x, y and z (denoted as the real space for the sake of simplification) and $\hat{X}$, $\hat{Y}$ and $\hat{Z}$ denote the new real coordinate system of the computational space, i.e., the bluish area of Fig. 1(b). Let us further denote by **L** the 3x3 diagonal matrix

$$L = \begin{bmatrix} X' & 0 & 0 \\ 0 & Y' & 0 \\ 0 & 0 & Z' \end{bmatrix},$$ where X' = dX/dx, Y' = dY/dy and Z' = dZ/dz. In our implementation [4], X', Y' and Z'

are simply piecewise-constant functions equal to 1 in $\Omega_1$ and equal to $f_{PML}$ (the complex PML coefficient) in the PML region $\Omega_2$ with a finite thickness $d_{PML}$. A classical result of PML theory [3] is that, if **E** and **H** are electromagnetic fields that satisfy Maxwell's equations in the (x,y,z) Cartesian coordinate system (real space) for a given frequency (eventually complex), the new electromagnetic fields

$$\hat{\mathbf{H}} = \mathbf{L}^{-1}\mathbf{H}, \ \hat{\mathbf{E}} = \mathbf{L}^{-1}\mathbf{E}, \tag{S7}$$

also satisfy the same Maxwell's equations in the new coordinate system provided that one renormalizes the permeability and permittivity tensors, according to

$$\hat{\boldsymbol{\mu}} = \mathbf{L}\boldsymbol{\mu}\mathbf{L}/\|\mathbf{L}\|, \ \hat{\boldsymbol{\varepsilon}} = \mathbf{L}\boldsymbol{\varepsilon}\mathbf{L}/\|\mathbf{L}\|, \tag{S8}$$

where $\|\circ\|$ denotes the determinant. It is easily found that $(\mathbf{E}\cdot\partial(\omega\boldsymbol{\varepsilon})/\partial\omega\mathbf{E} - \mathbf{H}\cdot\partial(\omega\boldsymbol{\mu})/\partial\omega\mathbf{H})d^3\mathbf{r}$ remains unchanged under the PML transform: *it is an invariant of the transform* that can be calculated with any PML. Indeed, using Eqs. (S7) and (S8), it is easily shown that $(\mathbf{E}\cdot\partial(\omega\boldsymbol{\varepsilon})/\partial\omega\mathbf{E} - \mathbf{H}\cdot\partial(\omega\boldsymbol{\mu})/\partial\omega\mathbf{H})$ $= \|\mathbf{L}\|(\hat{\mathbf{E}}\cdot\partial(\omega\hat{\boldsymbol{\varepsilon}})/\partial\omega\hat{\mathbf{E}} - \hat{\mathbf{H}}\cdot\partial(\omega\hat{\boldsymbol{\mu}})/\partial\omega\hat{\mathbf{H}})$, and since $d^3\mathbf{r} = dxdydz = d\hat{X}d\hat{Y}d\hat{Z}/\|\mathbf{L}\| = d^3\hat{\mathbf{r}}/\|\mathbf{L}\|$, we obtain

$$(\mathbf{E}\cdot\partial(\omega\boldsymbol{\varepsilon})/\partial\omega\mathbf{E} - \mathbf{H}\cdot\partial(\omega\boldsymbol{\mu})/\partial\omega\mathbf{H})d^3\mathbf{r} = (\hat{\mathbf{E}}\cdot\partial(\omega\hat{\boldsymbol{\varepsilon}})/\partial\omega\hat{\mathbf{E}} - \hat{\mathbf{H}}\cdot\partial(\omega\hat{\boldsymbol{\mu}})/\partial\omega\hat{\mathbf{H}})d^3\hat{\mathbf{r}}, \tag{S9}$$

(Q.E.D).

Therefore, the mode volume *V* defined in Eq. (9) (main text) can be numerically calculated virtually with any numerical tool. It is a powerful consequence of the present formalism, which is due to the fact that the important quantity *V is an invariant of coordinate transforms in Maxwell's equations*. Note that the presence of unconjugated products of the form $\mathbf{E}\cdot\partial(\omega\boldsymbol{\varepsilon})/\partial\omega\mathbf{E}$ is crucial to fulfill this property; the usual mode volume defined in Eq. (2) (main text) *is not* an invariant under complex coordinate transforms. In practice, provided that the complex coefficient and the thickness of the PML are large enough to warrant vanishing fields on the boarder of the computational domain, any PML can be used. Indeed, we have checked in all our examples that the calculated volumes are independent of the numerical parameters $d_{PML}$ and $f_{PML}$. Moreover, Table S1 illustrates the invariance of the integral in Eq. (S5) with respect to the size of the domain $\Omega_1$ for the fundamental quasi-normal mode of a gold sphere of radius r = 100 nm in air. The domain $\Omega_1$ is a sphere of radius R and three different sizes of $\Omega_1$ are considered, R=0.15, 1 and 2 µm. The integral of Eq. (S5) can be separated into two parts. The first part $I_1$ corresponds to the integration in a real spherical volume of radius R (the domain $\Omega_1$), and the second part $I_2$ corresponds to an integration in the PML region (domain $\Omega_2$). From the Table, we observe that, as the size of $\Omega_1$ increases, the integral over $\Omega_1$ increases because of the QNM divergence, but this increase is exactly compensated by an increase (with the opposite sign) of the integral over $\Omega_2$, so that the integral over $\Omega_1 \cup \Omega_2$ remains constant and equal to 1 with a 9-digits accuracy. Note that N is not strictly equal to 1 as it should be in theory only because of the numerical evaluation of the integral.

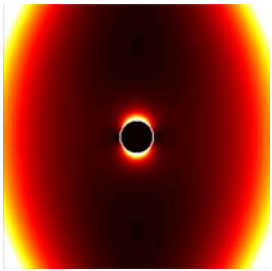

| R (µm) | $I_1$ (Integral over $\Omega_1$) | $I_2$ (Integral over $\Omega_2$) | N = $I_1$ + $I_2$ |
|---|---|---|---|
| 0.15 | 0.6193619 - 0.4489967i | 0.3806381 + 0.4489967i | 1 |
| 1 | 6.5664192 +0.4912745i | -5.5664192 – 0.4912745i | 1 |
| 2 | 1052.2978 – 1235.2268i | -1051.2978 + 1235.2268i | 1 |

**Table S1. Invariance of the mode volume with respect to the size of the domain $\Omega_1$.** We have calculated the integral N in Eq. (S5) for different sizes of the domain that is left without PML, the domain $\Omega_1$ (a sphere of radius R). The calculation holds for a gold nanosphere of radius r = 100 nm in air, for which the complex frequency of the fundamental (dipole-like) QNM is $2\pi c/\tilde{\omega} = 0.607 + 0.239i$ µm (the left inset shows a map of the QNM intensity $|E|^2$). The same Drude model as in the main text has been used for the permittivity of gold. As expected, the value of the integral is N = 1 (the mode has been normalized by the value of N for R = 2 µm) whatever the size of the domain $\Omega_1$. Both integrals $I_1$ and $I_2$ increase as R increases but they exactly compensate each other to keep the total integral constant.

To calculate the QNMs in the presence of a PML, we use a frequency-domain method known as the aperiodic Fourier Modal Method (a-FMM) [5], implemented either in Cartesian or in cylindrical coordinates [6]. Any other frequency-domain fully-vectorial method such as Finite Element Methods can be used. The QNM computation consists in looking for the complex poles of any scattering coefficient associated to the geometry under study. This is performed by solving Maxwell's equations for complex frequencies with an iterative algorithm. After the pole is found, the QNM fields $\tilde{\mathbf{E}}_n$ and $\tilde{\mathbf{H}}_n$ are calculated for a complex frequency $\tilde{\omega}$ close to $\tilde{\omega}_n$ (to avoid the singularity at $\tilde{\omega} = \tilde{\omega}_n$) inside and outside the PMLs and the mode volume is deduced by a numerical integration of this field, see Eq. (9).

## 3. Additional numerical tests

### 3.1. Coupled photonic-crystal microcavities

The aim of this Section is to test the present formalism with a "simple" example, namely two coupled photonic-crystal microcavities. Neither absorption nor material dispersion is considered in the system initially investigated in [7], but the physics associated to the dipole-field coupling is very rich and allows us to illustrate all the strength of a definition of a complex mode volume.

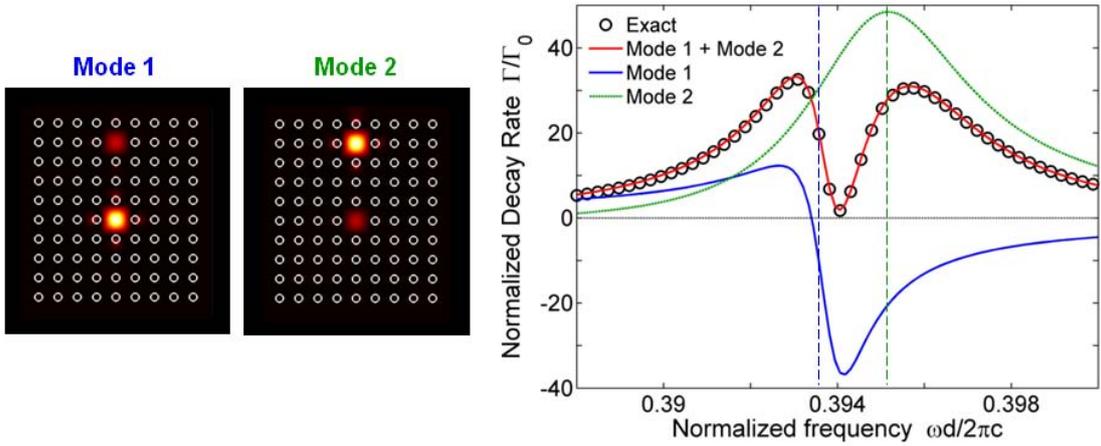

**Figure S1. Purcell factor of a coupled two-mode system.** Two cavities are formed by removing two rods in a finite-size photonic-crystal formed by a 9 × 10 array of semiconductor rods (refractive index of 3) in air. For an electric field polarized parallel to the rods, the system supports two dominant modes whose electric-field intensity is shown in the side panels. Their normalized complex frequencies are $\tilde{\omega}_1 a/2\pi c = 0.3938 – 0.0006i$ ($Q_1 = 300$) and $\tilde{\omega}_2 a/2\pi c = 0.3949 – 0.0023i$ ($Q_2 = 90$), with $a$ the photonic-crystal period and $c$ the speed of light. The real part of the complex eigenfrequencies are represented by vertical dashed lines. The spontaneous emission rate of an emitter located in the upper cavity with an electric dipole parallel to the rods is given by the black circles. With the present formalism, the total emission rate is given by the sum of the independent contributions $\Gamma_1$ and $\Gamma_2$ of the two modes, which are calculated analytically with Eq. (8) and shown by the solid blue and dashed green curves, respectively. Their sum (bold red curve) predicts a total emission rate in quantitative agreement with fully-vectorial calculations. In particular, the dip is precisely predicted; it comes from the peculiar contribution of mode 1 to the total emission, which highlights the dominant contribution of the imaginary part of its mode volume.

The system consists of two microcavities that are formed by removing two rods in a 2D photonic crystal made of a square array of semiconductor rods in air, see Fig. S1. Because of the proximity to the photonic-crystal boundary, the upper cavity (labeled 2) has a low $Q$-factor limited by leakage into the air clad, whereas the bottom cavity (labeled 1) has a higher $Q$-factor. We consider the spontaneous emission rate of a dipole polarized parallel to the rods and located in the center of the upper cavity. The spontaneous emission rate (black circles in Fig. S1), calculated with a high accuracy with a fully-vectorial aperiodic Fourier modal method (a-FMM) [5], exhibits the signature of a Fano-like resonance. To understand the dip in the spectrum, it is convenient to assume that the bottom cavity is absent. Then we face the simple problem of dipole emission into a single cavity with a low quality factor, for which one obtains a classical Lorentzian-shape spectrum (not shown) centered around a normalized frequency of 0.394. When adding the second cavity, the bottom mirror of the top cavity is modified. Intuitively, the reflectance modulus remains roughly the same, but as one scans over the resonance of the bottom cavity, the phase of the reflection coefficient varies. When the reflected wave is out of phase, destructive interference occurs, explaining the appearance of a narrow dip with a spectral width approximately equal to the linewidth of the high-Q bottom cavity.

We have calculated the QNMs of the coupled cavities with the a-FMM. The electric-field-intensity distributions are superimposed on the geometry in Fig. S1. Mode 1 is localized in the bottom cavity and has the higher quality factor ($Q_1 = 300$), while mode 2 is dominantly localized in the top cavity ($Q_2 = 90$). Because the materials are non-dispersive, the QNMs are orthogonal (see Section 1). Thus the off-diagonal terms in Eq. (5) are strictly equal to zero and the linear system of equations is diagonal and can be solved analytically. The normalized spontaneous emission rate can be rigorously written as a sum of independent contributions, $\Gamma/\Gamma_0 = \Gamma_1/\Gamma_0 + \Gamma_2/\Gamma_0$, $\Gamma_1$ and $\Gamma_2$ being given by Eqs. (8)-(10) in the main text. The solid blue and dashed green curves in Fig. S1 respectively represent the contributions $\Gamma_1/\Gamma_0$ and $\Gamma_2/\Gamma_0$, and the bold red curve is the sum.

Several comments are worth being mentioned. As expected, the dashed green curve that corresponds to a direct coupling between a cavity mode and an atom placed inside the cavity has a quasi-Lorentzian shape. In contrast, the solid blue curve, obtained for the atom located away from the maximum of the mode field distribution, is much more interesting. Remarkably, the contribution of mode 1 to the total decay rate may be either positive or negative and is not Lorentzian at all. In our opinion, this numerical result constitutes a strong evidence that dipole/cavity-mode coupling cannot be described with the usual $Q/V$ formula of Eq. (1) with a real and positive volume. Finally, it is also remarkable to see how accurate the present formalism is; the absolute deviation between the red curve and the black circles does not exceed 0.025 over the whole spectral range.

### 3.2. Plasmonic nanocylinder cavity

In addition to coupled photonic-crystal microcavities and metallic nanorod antennas, we have also checked the reliability of the generalized Purcell formula derived in the main text [see Eq. (8)] by applying it to a plasmonic nanocylinder cavity, see Fig. S2. This nanocavity has been recently designed for generating polarization entangled photons with a good fidelity from the bi-exciton cascade of a single quantum dot [8]. Similar metal-coated nanocylinders have also attracted much attention recently for realizing nanolasers [9].

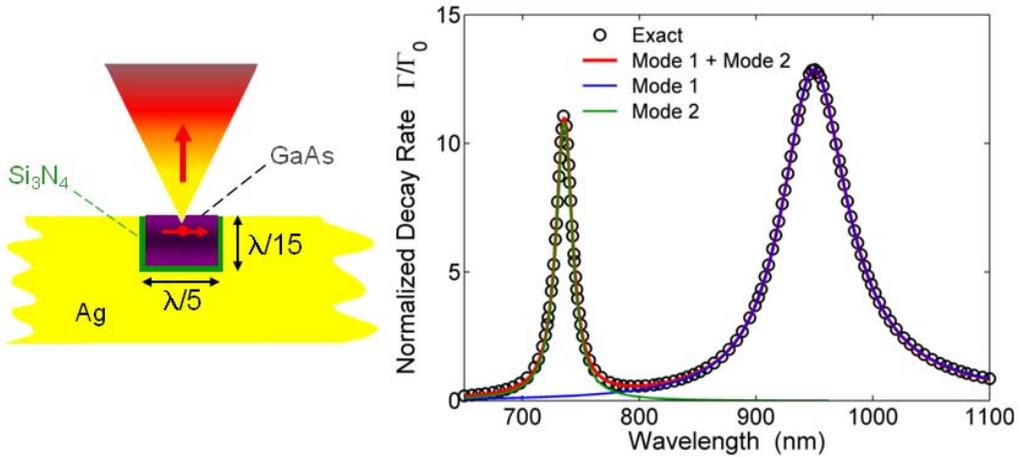

**Figure S2. Plasmonic nanocylinder cavity.** The resonator is a GaAs nanocylinder (refractive index $n = 3.45$, radius $R = 100$ nm and length $L = 65$ nm) coated with silver. A thin dielectric $Si_3N_4$ adlayer (thickness $e = 5$ nm) is added between metal and semiconductor. A quantum dot is located on-axis at a 8-nm distance from the GaAs/air interface. The cavity supports two dominant modes in the spectral range of interest, $2\pi c/\tilde{\omega}_1 = 950 + 34i$ nm and $2\pi c/\tilde{\omega}_2 = 735 + 8i$ nm. The spontaneous decay rate of the quantum dot has been calculated with the a-FMM implemented in cylindrical coordinates [6]. The fully-vectorial data (black circles) are in excellent agreement with the present formalism (bold red curve), whose predictions are the sum of the contributions of the two modes. These independent contributions are calculated with Eq. (8) and they are shown by the solid blue and dashed green curves.

Circles in Fig. S2 show the decay rate spectrum for a quantum emitter placed on-axis at a 8-nm distance from the cylinder aperture. The data are calculated with the a-FMM implemented in cylindrical coordinates [6]. In comparison with photonic-crystal microcavities, the mode lifetime is much shorter ($Q \approx 15$ is limited by Ohmic losses incurred in the silver clad), but since the mode confinement is much tighter ($V \approx 0.002\lambda^3$), the spontaneous decay rate is comparable; a 12-fold acceleration compared to the bulk emission is achieved for $\lambda = 950$ nm. Note that this plasmonic nanocylinder geometry has been recently implemented with nitrogen-vacancy color centers emitting in diamond [10] and that the experimental results largely confirm the main trends, a large bandwidth of 30 nm ($Q = 20$ at $\lambda = 600$ nm) and a spontaneous decay acceleration of 6.

With the metal-coated nanocylinder cavity, we are facing two independent resonances at $\lambda = 950$ and 735 nm, which result from the bouncing back and forth of two different plasmonic modes of the metal-coated semiconductor cylinder. To calculate the associated QNMs, one needs an analytical continuation of the permittivity $\varepsilon_{Ag}$ of silver for complex frequencies. For that purpose, we have taken tabulated data for $\varepsilon_{Ag}$ in [11], we have fitted them by a Drude model, $\varepsilon_{Ag} = \varepsilon_\infty - \omega_p^2/(\omega^2 + i\omega\gamma)$ with $\varepsilon_\infty = 3.7$, $\omega_p = 1.33\times10^{16}$ s$^{-1}$ and $\gamma = 1.06\times10^{14}$ s$^{-1}$, and have used the fitted expression for calculating $\varepsilon_{Ag}$ for complex frequencies. The solid blue and dashed green curves are independently obtained with the single-mode-approximation of our formalism, see Eq. (8), and their sum is given by the bold red curve. The agreement is remarkable, since the resonance linewidths and the small values of the decay rate are in quantitative agreement.

### 3.3. Detailed analysis of the limitations

The formalism presented in the main text aims at describing the coupling between an emitter and an optical resonance. Therefore, we have assumed that the electromagnetic field radiated by the source can be accurately calculated by a few QNMs, see Eq. (4) in the main text. In order to analyze the limitations of this assumption, we have calculated the spontaneous decay rate of the system in Fig. 2 as a function of the emitter-nanorod distance $d$. The results are presented in Fig. S3. Circles and squares correspond to fully-vectorial calculations of the decay rate for an emission frequency matched to the resonance frequency and for on-axis (red arrow) and off-axis (blue arrow) dipoles. They are in excellent agreement with the predictions (colored solid and dashed curves) obtained with the generalized Purcell formula given by Eq. (8); the decay rates are predicted with a high accuracy for values ranging over two orders of magnitude.

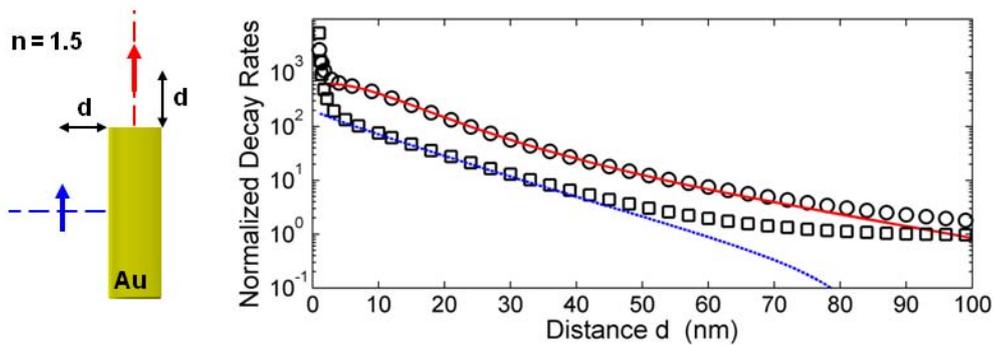

**Figure S3. Single metallic nanorod, decay rate as a function of the emitter-nanorod distance.** The gold nanorod is the same as in Fig. 2 (main text). We have calculated the total decay rate as a function of the emitter-nanorod distance $d$ for an emission frequency matched to the resonance frequency ($\lambda = 920$ nm). Circles are fully-vectorial data and the colored curves are obtained with the generalized Purcell formula [Eq. (8) in the main text]. The solid red and dashed blue curves correspond to a dipole oriented parallel to the nanorod and located on-axis (red arrow) and off-axis (blue arrow), respectively. For short and long distances, the coupling to the nanorod resonance is not dominant; the decay rate is dominated either by quenching in the metal or by direct coupling to free-space.

The limitations of the present formalism are rather restricted. Deviations are only observed for small and large distances. Indeed, neither the sudden increase of the local density of states due to a direct electrostatic coupling with the metal (quenching) nor the direct coupling to free-space that becomes dominant as $d$ increases is included in our assumption that the emitter is mainly coupled to a few modes of the resonator. For small distances, the decay rate becomes dominated by the non-resonant excitation of higher-order modes [12]. On the other hand, for large distances, the decay rate is dominated by the coupling to the free-space continuum. Note however that this contribution can be handled analytically and added *a posteriori* in Eq. (8).

The radiation continuum associated to a direct emission in free-space can be taken into account by adding 1 in the expression of the normalized decay rate, $\Gamma/\Gamma_0 = 1 + F$, rendering the difference between the markers (circles and squares) and the colored curves in Fig. S3 invisible for large distances.

More generally, the limitations of expanding the electromagnetic field onto a small set of discrete modes [Eq. (4) in the main text] is a mathematical issue linked to the completeness of the QNMs. This question has been addressed in the 90's in the case of simple resonator geometries; it has been demonstrated that the QNMs form a complete basis for describing the field *inside* an open dielectric one-dimensional microcavity [13]. However, there does not exist, to our knowledge, any general result on the completeness of the QNMs for three-dimensional complex geometries. In particular for the case of plasmonic nanostructures for which we are interested in calculating the field *outside* the particle.

Finally, we would like to emphasize that the motivation of the present work is not to calculate the exact value of the total spontaneous decay rate (that would probably require the calculation of many modes), but rather to have a physically-sound and mathematically-safe decomposition into a small number of modes that are expected to all have a deep physical meaning.